\documentclass{aa}
\usepackage{graphicx}
\usepackage{color}
\usepackage{soul}
\usepackage{float}
\usepackage{upgreek}
\usepackage{txfonts}
\usepackage[colorlinks=true,citecolor=blue,linkcolor=blue,urlcolor=blue]{hyperref}
\usepackage{natbib,twoopt}
\bibpunct{(}{)}{;}{a}{}{,} 
\makeatletter
\newcommandtwoopt{\citeads}[3][][]{\href{http://adsabs.harvard.edu/abs/#3}%
{\def\hyper@linkstart##1##2{}%
\let\hyper@linkend\@empty\citealp[#1][#2]{#3}}}
\newcommandtwoopt{\citepads}[3][][]{\href{http://adsabs.harvard.edu/abs/#3}%
{\def\hyper@linkstart##1##2{}%
\let\hyper@linkend\@empty\citep[#1][#2]{#3}}}
\newcommandtwoopt{\citetads}[3][][]{\href{http://adsabs.harvard.edu/abs/#3}%
{\def\hyper@linkstart##1##2{}%
\let\hyper@linkend\@empty\citet[#1][#2]{#3}}}
\newcommandtwoopt{\citeyearads}[3][][]%
{\href{http://adsabs.harvard.edu/abs/#3}
{\def\hyper@linkstart##1##2{}%
\let\hyper@linkend\@empty\citeyear[#1][#2]{#3}}}
\makeatother

\begin{document}

   \title{Potassium detection in the clear atmosphere of a hot-Jupiter}
	\titlerunning{WASP-17b transmission spectroscopy}
   \subtitle{FORS2 transmission spectroscopy of WASP-17b}

   \author{E. Sedaghati\inst{1,2,3}
   			\and
            H. M. J. Boffin\inst{1,4}
            \and
            T. Je\v{r}abkov\'{a}\inst{5,6}
            \and
			A. Garc\'{i}a Mu\~{n}oz\inst{3}
			\and
			J. L. Grenfell\inst{2,3}        
			\and            
			A. Smette\inst{1}
            \and
			V. D. Ivanov\inst{1,4}
			\and            
            Sz. Csizmadia\inst{2}
            \and
            J. Cabrera\inst{2}
			\and
			P. Kabath\inst{6}
            \and
            M. Rocchetto\inst{7}
            \and
            H. Rauer\inst{2,3}
            }

   \institute{
			European Southern Observatory, Alonso de C\'ordova 3107, Casilla 19001, Santiago, Chile\\
              \email{[esedagha@eso.org; hboffin@eso.org]}
         	\and
            Institut f\"ur Planetenforschung, Deutsches Zentrum f\"ur Luft- und Raumfahrt, Rutherfordstr. 2, 12489 Berlin, Germany
            \and
            Zentrum f\"ur Astronomie und Astrophysik, Technische Universit\"at Berlin, Hardenbergstra$\upbeta$e 36, 10623 Berlin, Germany
			\and
            European Southern Observatory, Karl-Schwarzschild-Str. 2, 85748 Garching bei M\"unchen, Germany            
            \and
            Astronomical Institute, Charles University in Prague, Vhole\v{s}ovi\v{c}k\'ach 2, CZ-180 00 Praha 8, Czech Republic
            \and
            Astronomical Institute ASCR, Fri\v{c}ova 298, Ond\v{r}ejov, Czech Republic
			\and 
            Department of Physics \& Astronomy, University College London, Gower Street, London, WC1E 6BT, UK
            }

   \date{Received 10 June 2016 ; accepted 14 September 2016}

 \abstract{We present FORS2 (attached to ESO's Very Large Telescope) observations of the exoplanet WASP-17b during its primary transit, for the purpose of differential spectrophotometry analysis. We use the instrument in its Mask eXchange Unit (MXU) mode to simultaneously obtain low resolution spectra of the planet hosting star, as well as several reference stars in the field of view. The integration of these spectra within broadband and smaller 100\AA~bins provides us with ‘white' and spectrophotometric light curves, from 5700 to 8000\AA. Through modelling the white light curve, we obtain refined bulk and transit parameters of the planet, as well as wavelength-dependent variations of the planetary radius from smaller spectral bins through which the transmission spectrum is obtained. The inference of transit parameters, as well as the noise statistics, is performed using a Gaussian Process model. We achieve a typical precision in the transit depth of a few hundred parts per million from various transit light curves. From the transmission spectra we rule out a flat spectrum at >3$\sigma$ and detect marginal presence of the pressure--broadened sodium wings.  Furthermore, we detect the wing of the potassium absorption line in the upper atmosphere of the planet with 3-$\sigma$ confidence, both facts pointing to a relatively shallow temperature gradient of the atmosphere.  These conclusions are mostly consistent with previous studies of this exo--atmosphere, although previous potassium measurements have been inconclusive.}


   \maketitle
%

\section{Introduction}

In the short number of years in which extrasolar planetary systems have been detected and studied, there has been great progress made in understanding these alien worlds and characterising their intrinsic properties.  Methods such as radial velocity and transit monitoring, have facilitated the measurement of bulk densities, as well as some initial approximations of the structure and atmospheric properties of extrasolar planets.  It is the latter with which we are concerned here, whereby spectroscopic observations of exoplanetary primary transits lead to setting constraints on the chemical composition and physical processes within exo-atmospheres.

Through transmission spectroscopy one analyses the wavelength dependent variations of the exoplanetary radius.  Such minute variations are caused by the presence of an optically thick atmosphere at specific wavelengths, dictated by the absorption and scattering characteristics of the gas and aerosols present near the planet's terminator.  These wavelength dependent opacity variations are scanned across the spectrum, and used to probe chemical compositions of the planets' atmospheres \citep{Seager2000,Brown2001}.  The ideal candidates for such studies are those planets with extended atmospheres, i.e. with large atmospheric scale heights, such as the so-called hot Jupiters  \citep{Seager1998}. The search for signatures of exo-atmospheres began as early as the time when predictions were being made.  For instance \cite{Rauer2000} and \cite{Harris2012}, among others, looked for spectral signatures due to absorption by extended atmospheres surrounding 51 Peg b and $\tau$ Bo\"otis Ab.

Traditionally, transmission spectroscopy has been dominated by space-based facilities, due to their obvious advantage of not being affected by telluric extinction \citep{Charbonneau2002,Ehrenreich2007,Pont2008,Gibson2012a,Deming2013,Knutson2014,Swain2014}.  However, large ground-based telescopes, with instruments meeting the requirements of transmission spectroscopy science goals, VLT/FORS \citep{Appenzeller1998}, Gemini/GMOS \citep{Hook2004} or Magellan/IMACS \citep{Bigelow1998} being the most effective of those, have also been able to contribute to such studies.  \cite{Gibson2013a,Gibson2013b} among others have shown the capabilities of GMOS in producing transmission spectra for a number of transiting hot Jupiters.  With its MOS (Multi Object Spectroscopy) and MXU (Mask eXchange Unit) modes, the FORS2 (FOcal Reducer and low dispersion Spectrograph) instrument on the Unit Telescope 1 (UT1 -- Antu) of ESO's Very Large Telescope (VLT) has also been a key player in producing such spectra.  \cite{Bean2010,Bean2011} were able to obtain pioneering results for the GJ1214b transmission spectrum, an exoplanet in the mini Neptune regime, from the visible to NIR bands.  However, subsequent attempts at transmission spectroscopy analysis with this instrument were mostly unsuccessful due to systematics present in the data associated with prisms of the Longitudinal Atmospheric Dispersion Corrector (LADC).  Therefore, a project was started at ESO Paranal to address such issues \citep{Boffin2015}, by removing the inhomogeniously degrading coating from the prisms of the decommissioned FORS1's LADC, replacing them with their FORS2 counterpart.  The improvements were subsequently highlighted through transit observations of WASP-19b, the results of which were presented in \citet{Sedaghati2015}.

An essential issue plaguing all the aforementioned instruments, has been the role of instrumental systematics and their manifestation in the final transit light curves. As transmission spectroscopy heavily relies on extremely precise measurement of the transit depth (a few hundreds of ppm\footnote{Parts per million.} for hot Jupiters, down to a few ppm for terrestrial planets), and subsequently the relative planetary radius, consideration of such factors is crucial in determining the correct scaled radius values, as well as pragmatic estimation of their error budgets.  This fact is of paramount importance when a transmission spectrum is utilised together with atmospheric models in determining the physical properties of an exo-atmosphere.  The time-dependent correlated noise\footnote{Sometimes also refered to as `red' or systematic noise.} due to systematics is somewhat reduced through differential spectroscopy techniques employed in this work, however some effects are still expected to remain which originate from poorly understood sources.  In this paper we utilise a method with this apparent correlated noise component in mind, to model and analyse transit light curves. The method is based on the Gaussian process (GP) of \citet{Gibson2012b}, adapted for modelling time-correlated noise, which has been shown to provide conservative and realistic error estimations \citep{Gibson2013b} as compared to other parametric methods such as the Wavelet decomposition techniques of \citet{CarterWinn2009}. This is an essential point to consider when detecting exoplanetary atmospheric features with transmission spectroscopy, since underestimating the precision of radius measurements can lead to false or inaccurate characterisation of the atmosphere.

Here we report FORS2 observations of WASP-17b in MXU mode.  WASP-17b \citep{Anderson2010} is an ultra low density Jupiter size planet on a $\sim$3.74d, possibly retrograde orbit around an 11.6V magnitude F6V star.  It has a mass and radius of 0.486$\pm$0.032 M$_\text{Jupiter}$ and 1.991$\pm$0.081 R$_\text{Jupiter}$ respectively \citep{Anderson2011}, making it an extremely bloated hot Jupiter at only 0.06 $\rho_\text{Jupiter}$.  This together with an equilibrium temperature of 1771$\pm$35 K \citep{Anderson2011}, mean a relatively large atmospheric scale height, making it an ideal candidate for transmission spectroscopy studies.

This paper is structured as follows. Section 2 highlights our observations and data reduction, in Section 3 we present detailed transit data analysis steps, in Section 4 we show the results of the analysis and discuss the atmospheric characteristics of WASP-17b , in Section 5 we briefly discuss our atmospheric results for this exoplanet and finally in Section 6 a brief summary of our conclusions is presented.

\section{FORS2 Observations}

\begin{figure}[t!]
\includegraphics[width=\linewidth]{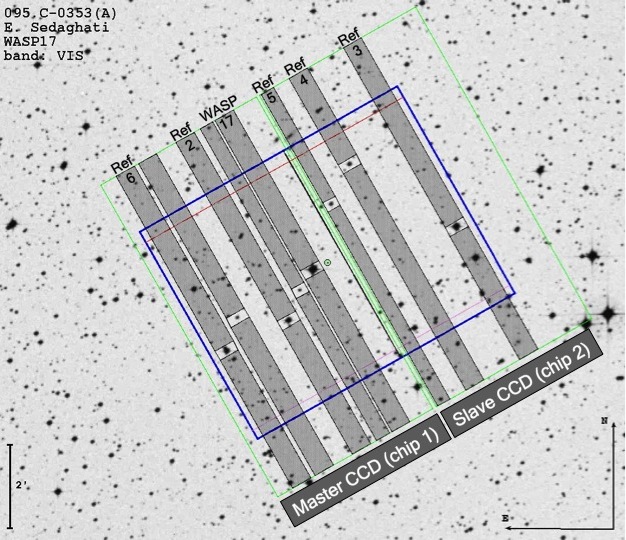}
\caption{A 10\arcmin $\times$10\arcmin ~plot encompassing the FoV of FORS2 (green box) and the area covered by the 2 (2k$\times$4k) MIT chip mosaic (in blue). The exact design and location of the MXU slits, with the relevant target designations are also shown. The grey regions are the sections of the CCD used for recording the individual stellar spectra.}
\label{fig:mask}
\end{figure}

A single transit of WASP-17b was observed using the 8.2m UT1 of the VLT with the FORS2 instrument on the 18th of June 2015, with the data taken as part of the programme 095.C-0353(A) (PI: Sedaghati).  FORS2 has a $6.8' \times 6.8'$ field of view, including a detector system consisting of a mosaic of two 2k$\times$4k, red optimised, MIT CCDs, utilised for the observations, all of which are visible in Fig. \ref{fig:mask}.  We used FORS2 in the MXU\footnote{The means that a custom-designed mask is used instead of the movable arms (MOS).} mode to observe the target and a number of reference stars simultaneously for $\sim$8.1h, covering the 4.38h of complete transit plus 2.3h prior to ingress and 1.4h post egress.  Observational conditions were clear for parts of the night, with some sporadic cirrus affecting the observations in transit.  The LADC was left in park position during the observing sequence, with the two prisms fixed at their minimal separation distance of 30mm. At the start of the observations (23:20UT), the field was at airmass of 1.48, moving through the zenith at 00:23UT and finally settling to 2.31 for the last exposure at 07:45UT.

The observations were performed with the 600RI grism plus the GG435 order sorter filter, covering the wavelength range\footnote{Note that the exact wavelength coverage for the various stars in the slits varies modestly depending on their positions on the CCD.} of $\sim$5500 to 8300\AA.  The exposure times were 35s throughout the night and the data were binned (2 $\times$ 2) yielding a scale of $0.25''$ per pixel.  We used the non-standard 200kHz readout mode, enabling us to spend more time on target for a fixed cadence ($\sim$70 s), due to the reduced readout time and lower gain of this mode.  This procedure allowed for 477 exposures, with some time lost to resetting the field rotator, which we ensured was done pre-ingress.  We created a customised mask with $15''$ wide slits to minimise slit losses.  The slits were mostly $30''$ long, with the exception of 2 to avoid overlapping dispersion profiles.  The mask set up on the sky is shown in Fig. \ref{fig:mask}.  A copy of the mask was created for the purpose of wavelength calibration, with narrow 1$''$ slits centred on the science slits.  Bias, flatfield and arc images were taken, as part of the routine daytime calibration sequences, before and after the observations.  Details of an optimised observational strategy with FORS2 have been presented by \cite{Boffin2016}.

\begin{figure}
\includegraphics[width=\linewidth]{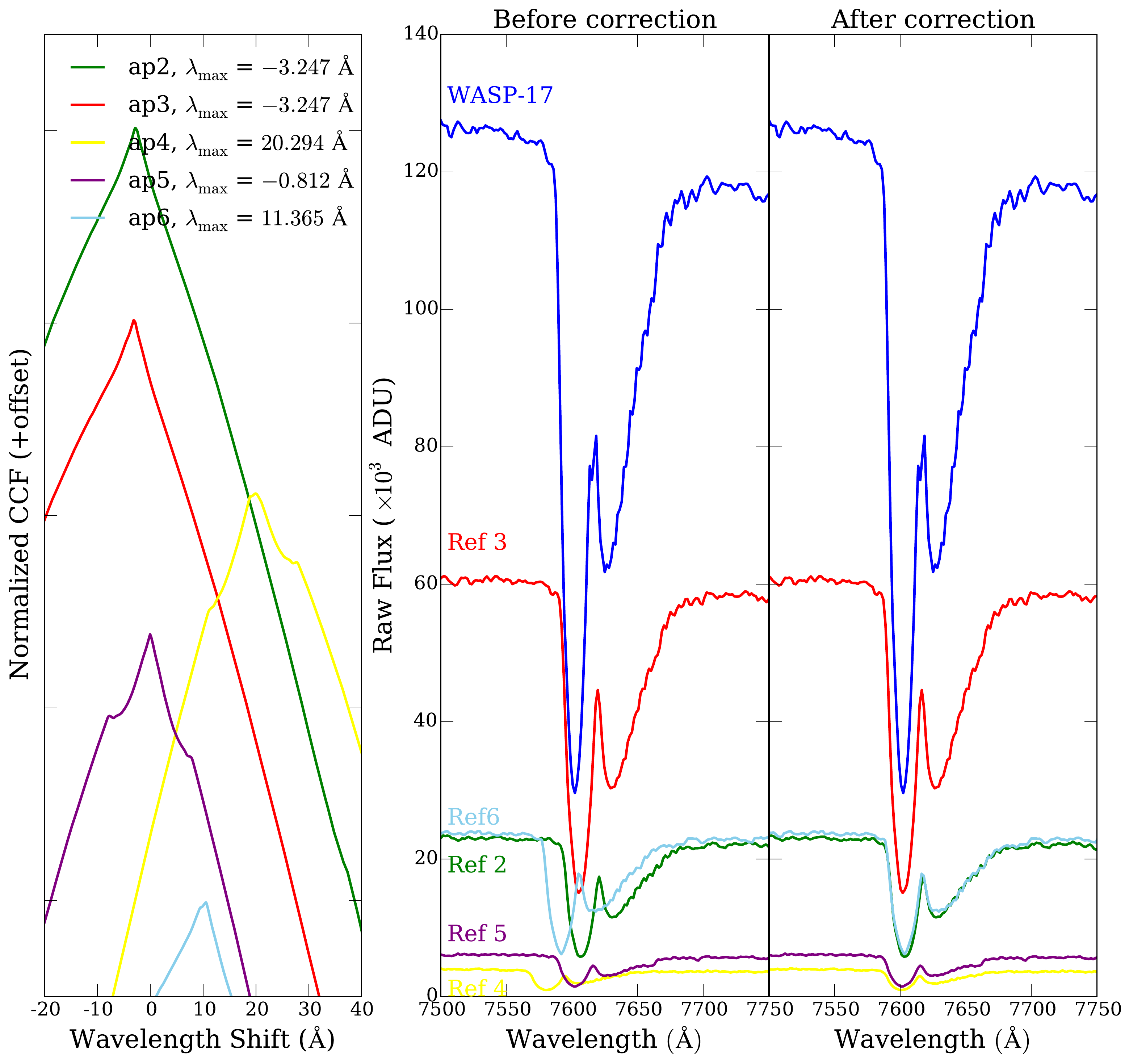}
\caption{\textit{(Left)} Cross correlation functions, calculated for the wavelength solutions of all reference stars, with respect to the target.  The maximising wavelength at each instance, presented in the figure, is then added to each wavelength solution for the correction of the spectra. \textit{(Right)} A zoom into the telluric $O_2$ absorption region of the spectra, before and after the wavelength correction procedure.}
\label{fig:CCF}
\end{figure}

\begin{figure}[t]
\includegraphics[width=\linewidth]{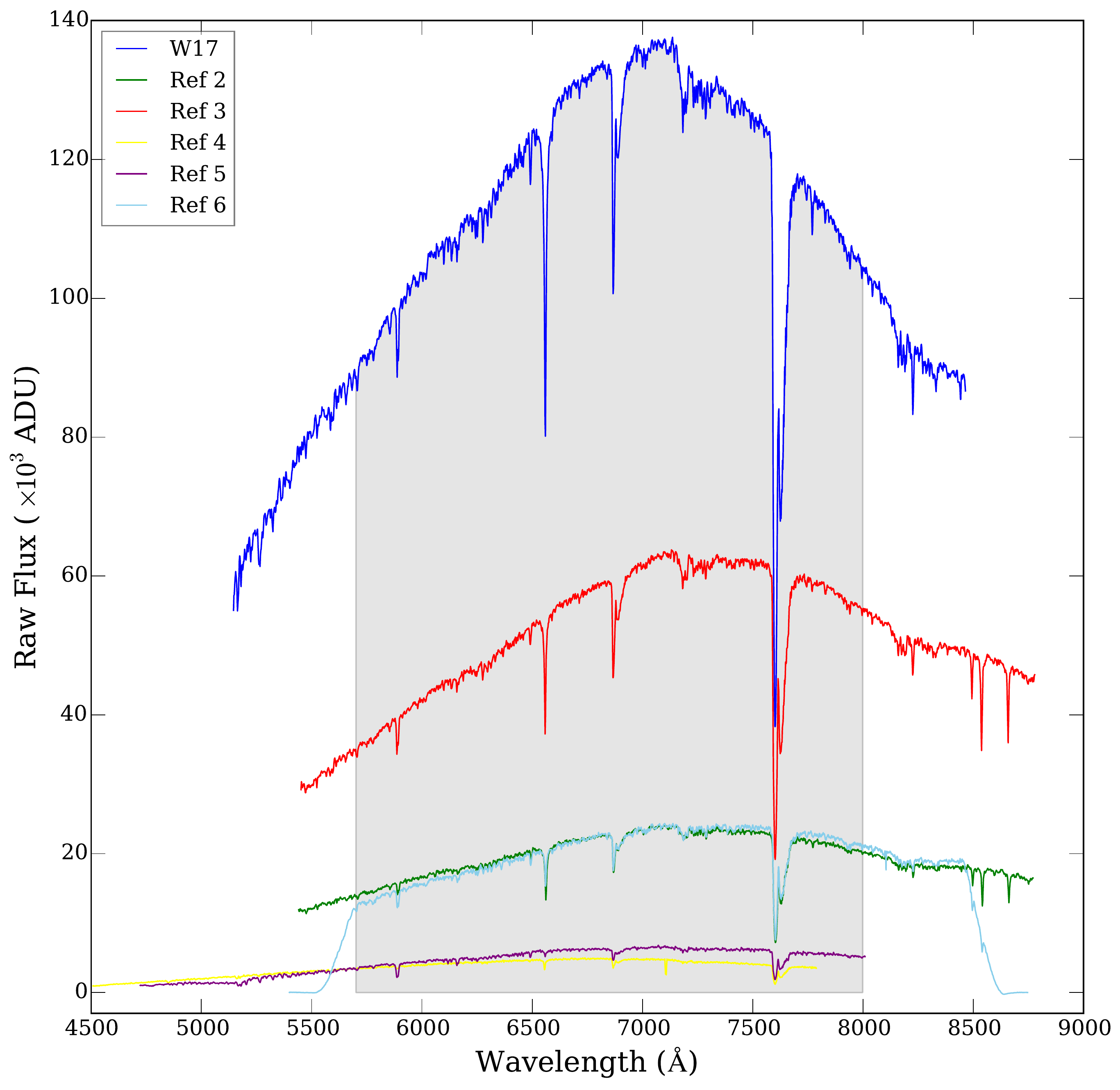}
\caption{Wavelength-corrected raw spectra of the target and reference stars, where the numbers are consistent with those in Fig. \ref{fig:mask}.  The grey shaded region represents the integration limits for the broadband light curve production.}
\label{fig:raw spec}
\end{figure}

\section{Data Reduction and Analysis}

We wrote a {\tt PyRAF}\footnote{This is a product of the Space Telescope Science Institute (STScI), for running {\tt IRAF} tasks based on the {\tt Python} scripting language.} pipeline for the reduction purposes of the data, optimising it to the specific science goals of this dataset.  The steps in the pipeline included the overscan and bias shape subtraction, spectral flat-fielding and wavelength calibration.  Subsequently, wavelength calibrated spectra were extracted for the target and 5 of the 7 reference stars, while simultaneously subtracting the sky background and removing any cosmic ray contamination.  There was a further need to finely tune the dispersion solution found for the reference stars due to imperfect centering of stars on the slits.  This was done by cross correlating the spectrum of each reference star with that of WASP-17, shown in Fig. \ref{fig:CCF}, to find a small shift between each pair.  The maximal solution was then added to the wavelength stamp of each reference star's dispersion relation.  This is a rather important point to consider, since differences in wavelength solutions of various stars would cause artificial features in the spectrophotometric light curves.  An example set of these extracted spectra is shown in Fig. \ref{fig:raw spec}.  A careful analysis of the choice of reference star and its impact on the final light curve is presented in the following section.

\subsection{Reference Star Analysis}

We observed a total of 8 targets through the custom designed mask, with wide slits placed on those targets.  Of the 7 reference stars, 5 were deemed bright enough for further analysis. We searched through a variety of catalogues for estimation of the spectral type of these stars, the results for which are given in Table \ref{tab:ref star}.  Furthermore, we also searched through a number of catalogs for variability in our reference stars.  From the Catalina survey \citep{Drake2009}, the star in slit 3 shows periodicity at $\sim$4.4 days at an amplitude of $\sim$0.1 mag. However, this result is rather tentative as the object is brighter than the 13 V$_{\text{mag}}$ saturation limit given by the survey.  Additionally, the duration of our observations is considerably shorter than the variability period and therefore any possible intrinsic variation should be insignificant for our purposes.

\begin{table}[t]
\centering
\caption{Spectral characterisation of the stars observed.}
\begin{tabular}{l c c c c}
\hline \hline
Star & UCAC4 ID$^a$ & $R_{\text{mag}}^{~~~b}$  & \textit{J}$-$\textit{K}$^c$ & Spectral Type\\
\hline
W17 & 310-085339 & 11.31 & 0.29$\pm$0.06 & F5.5$-$G8.0\\
Ref 2 & 310-085352 & 13.26 & 0.29$\pm$0.05 & F7.5$-$G8.0\\
Ref 3 & 310-085281 & 12.26 & 0.37$\pm$0.06 & G4.0$-$K1.0\\
Ref 4 & 310-085328 & 14.9 & 0.42$\pm$0.07 & G9.0$-$K3.0\\
Ref 5 & 310-085333 & 14.82 & 0.66$\pm$0.06 & K5.0$-$K7.0\\
Ref 6 & 310-085382 & 12.7 & 0.42$\pm$0.04 & G9.0$-$K2.0\\
\hline
\multicolumn{5}{l}{$^{a}$~{\scriptsize  The stellar catalogue IDs are given using references from \textit{The fourth US}}}\\
\multicolumn{5}{l}{~~{\scriptsize \textit{naval observatory CCD astrograph catalog} \citep[\textit{UCAC4;}][]{Zacharias2013}.}}\\
\multicolumn{5}{l}{$^{b}$~{\scriptsize \textit{R} magnitudes are obtained from the UCAC2 catalogue \citep{Zacharias2004}.}}\\
\multicolumn{5}{l}{$^{c}$~{\scriptsize \textit{J} and \textit{K} magnitude are obtained from the UCAC3 catalogue \citep{Zacharias2010}.}}\\
\multicolumn{5}{l}{~~{\scriptsize These magnitude differences are converted to spectral type by using estimates from}}\\ 
\multicolumn{5}{l}{~~{\scriptsize Table 2 of \citet{Ducati2001}, assuming all the stars are main sequence.}}
\end{tabular}
\label{tab:ref star}
\end{table}

\begin{figure*}[ht!]
\includegraphics[width=\textwidth,height=9cm]{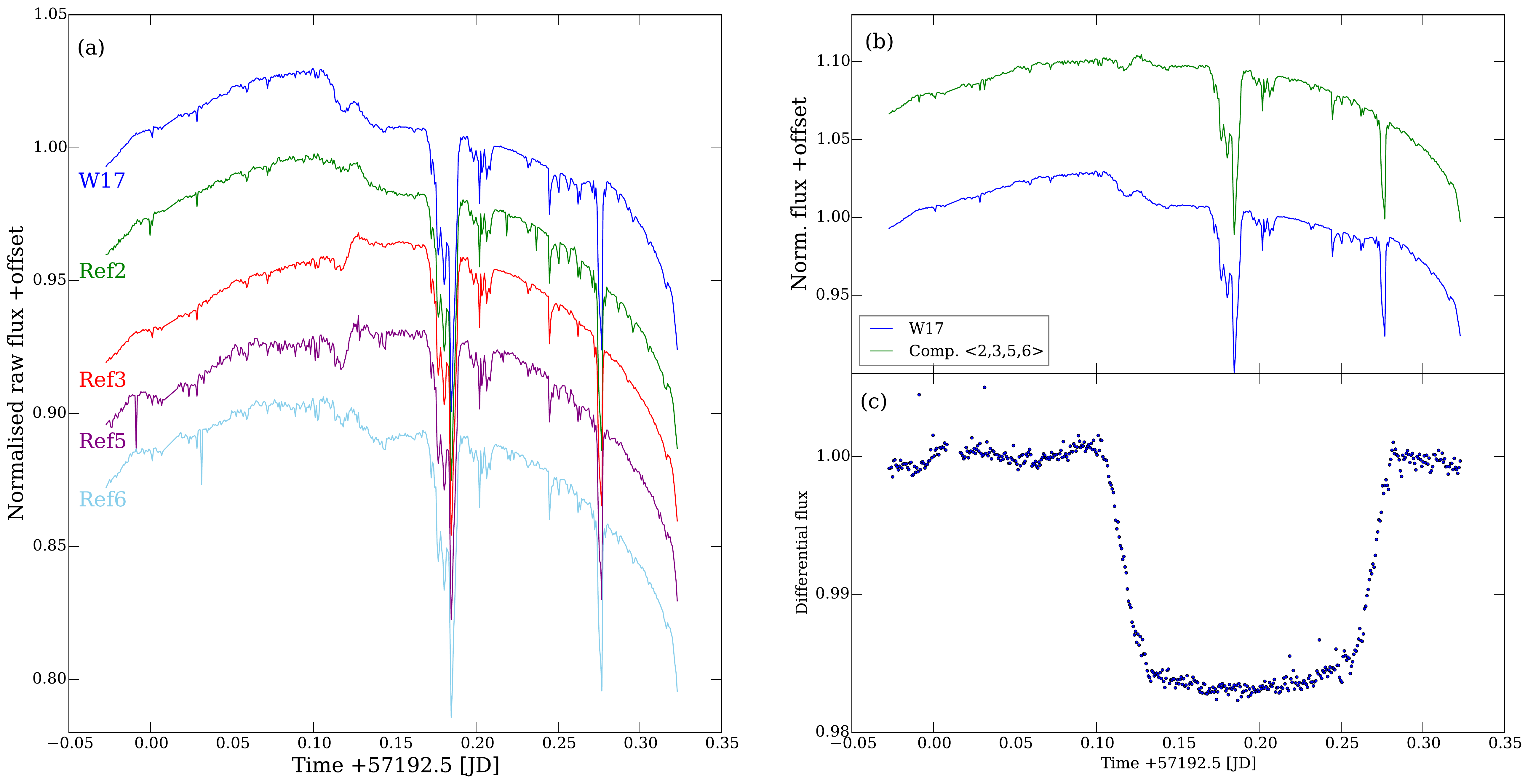}
\caption{\textit{(a)} Raw time-series light curves of the 6 stars observed and analysed with the custom designed mask, shown in Fig. \ref{fig:mask}. \textit{(b)} Raw light curves for WASP-17 and the composite comparison star made from the combination of all reference stars, excluding reference 4. \textit{(c)} normalised transit light curve for WASP-17b obtained through division of the two plots above and normalised to out-of-transit flux mean.  Light curves in panels \textit{a} and \textit{b} have been shifted for clarity.}
\label{fig:raw LCs}
\end{figure*}

We first integrate the individual spectra for the largest possible common wavelength domain of all 6 targets to produce the raw broadband light curves shown in Fig. \ref{fig:raw LCs}a, normalised to the mean flux and shifted for clarity.  We then produce differential light curves by simply dividing the normalised WASP-17 light curve by all the individual reference stars (Fig. \ref{fig:raw LCs}b), as well as all possible multiple-star combinations.  The final comparison star is created through the combination of all reference stars, excluding reference star 4, since its inclusion results in a transit light curve with larger systematics, as well as reducing the possible wavelength domain available for integration of spectra.  We also experimented with a composite comparison star that was obtained using a weighted mean of all the individual reference stars using the Signal to Noise Ratio (SNR) of individual transit light curves as the weights, however, this did not have any impact on the final transit light curve.  Hence, we create the final comparison using the arithmetic mean of the chosen reference stars (2, 3, 5, and 6).

\begin{figure}[t]
\includegraphics[width=\linewidth,height=8.845cm]{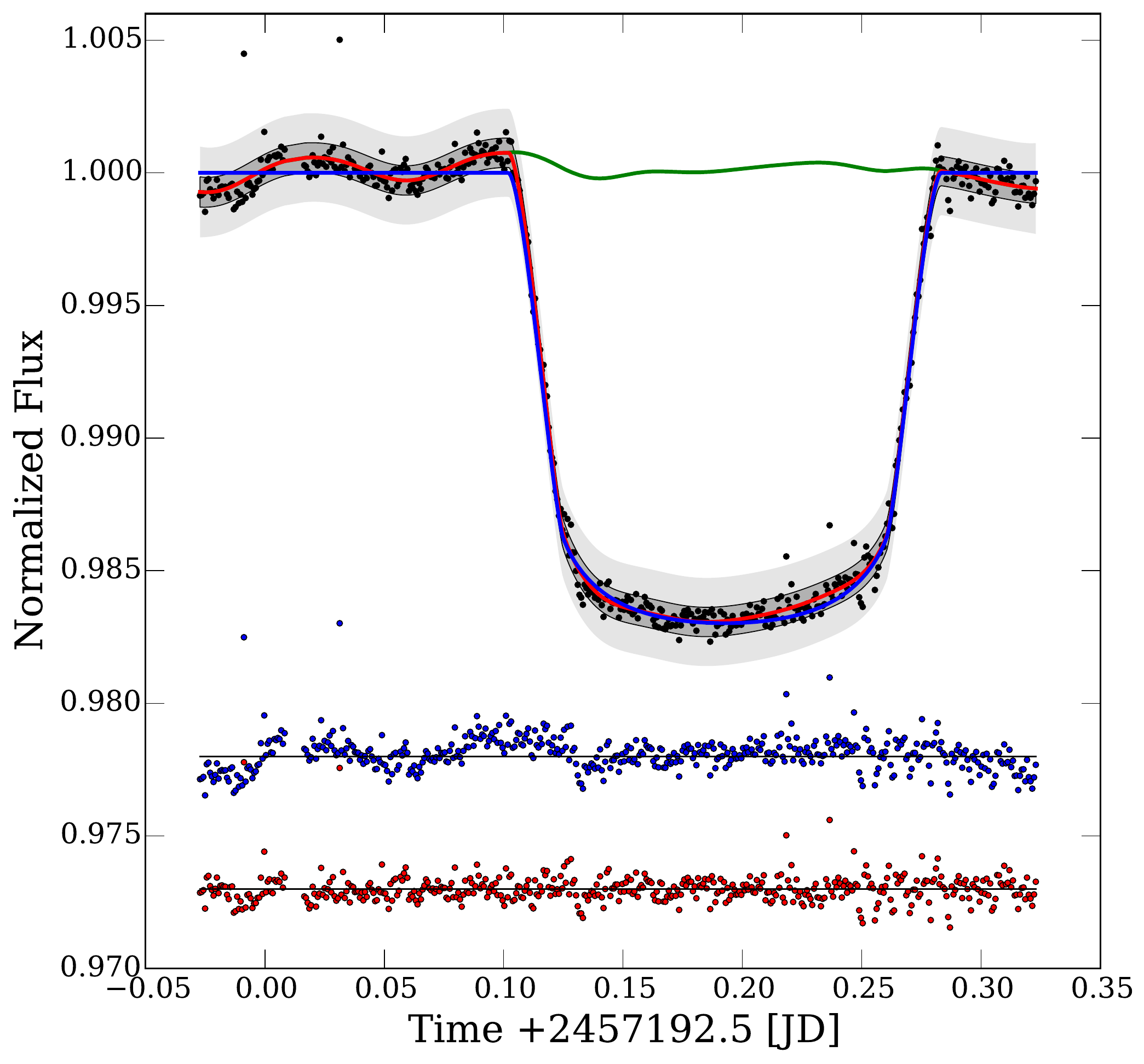}
\caption{Broadband (white) light curve for the transit of WASP-17b, fitted with a GP model described in §\ref{sec:GP model}. The best fitting transit light curve is shown in blue with the GP mean model given as solid red line. The residuals of both models are shown at the bottom, where they are shifted for clarity and their colours correspond to the models that they represent. The dark and light grey regions represent the 1- and 3-$\sigma$ predictions of the GP model, respectively. The green line shows the systematics model without the transit function.}
\label{fig:broadband_LC}
\end{figure}

\subsection{Spectral Statistical Analysis}

\begin{figure}
\includegraphics[width=\linewidth,height=8.72cm]{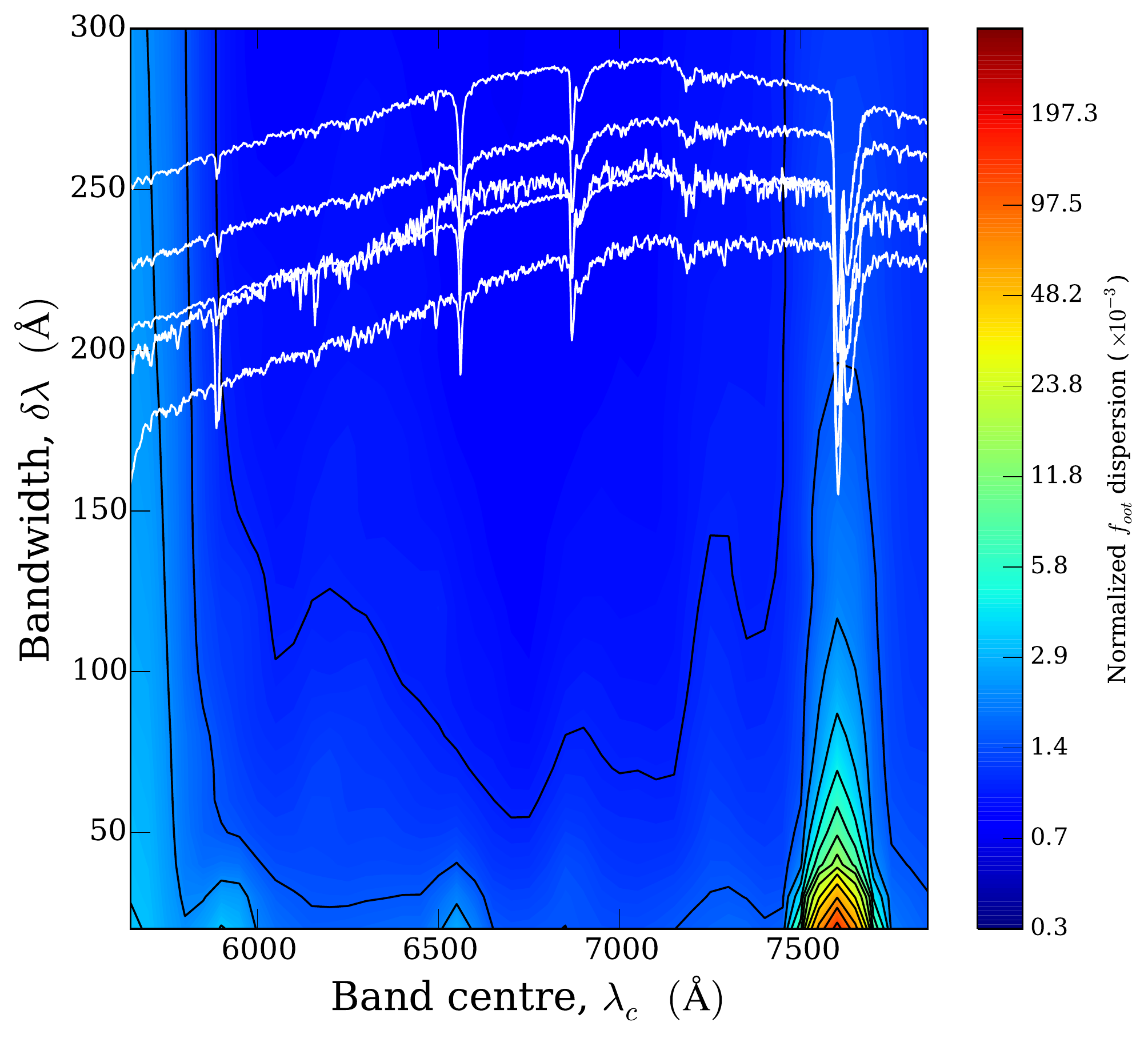}
\caption{Analysis of light curve precision across the entire wavelength domain, with the spectra of WASP-17 and all reference stars superimposed at the top.  Clearly, the regions containing telluric absorption features, require larger bandwidths in order to produce reasonable spectrophotometric light curves, for instance the O$_2$(A) absorption region at approximately 7600\AA.}
\label{fig:glob stat}
\end{figure}

In order to obtain the wavelength dependency of planetary radius, we need to produce spectrophotometric light curves.  This is done by integrating the obtained spectra within arbitrarily positioned filters of widths that are based on the quality of the light curves produced.  Typically, transmission spectra produced with FORS2 have resolutions of 200\AA~ \citep{Bean2010,Sedaghati2015}, with the exception of a few studies \citep[e.g.][]{Bean2011,Lendl2016} that produced spectra with 100\AA~bands.  The precisions of differential and detrended light curves are calculated for a range of bandwidths, measured across the entire wavelength domain.  This is done by calculating the dispersion of the normalised $\boldsymbol{\mathrm{f}}_{\text{oot}}$ for light curves produced at a central wavelength $\lambda_c$ and bandwidths\footnote{We start at 20\AA~minimum bandwidth, since below this integration size the algorithm that detects the transit locations automatically, runs into difficulty, due to the low SNR in the light curve.  Furthermore, for practical purposes, the calculations are performed with steps in bandwidth of 10\AA~to reduce computation time.} $\delta \lambda$ in the range of 20--300\AA.  This process is performed for the entire wavelength range, for $\lambda_c$ values spaced at 50\AA. Subsequently, to interpolate the results in between these steps, we apply a Gaussian convolution filter for a smoothed continuous dispersion map\footnote{Namely, this is done to obtain a continuous distribution of dispersion across the $\lambda_c$ vs. $\delta \lambda$ parameter space, with a two-dimensional Gaussian filter of $\sigma = 1$ for the kernel in both dimensions.}. This is essentially similar to the mean filter but having a different kernel, an application for which is given by \cite{Buades2005} for image denoising. The results from such statistical analysis of light curve precision across the entire spectrum is shown in Fig. \ref{fig:glob stat}.

It is quite evident that in certain regions a rather large bandwidth is required to obtain a precise light curve.  This is due to the presence of telluric absorption features that reduce the SNR in the produced light curve significantly.  Based on this analysis and taking into account the contours of Fig. \ref{fig:glob stat}, we choose 100\AA~as the bandwidth to be used for the production of the spectrophotometric light curves.  This value is chosen to provide us with high resolution transmission spectral points that are simultaneously precise enough for comparison with atmospheric models.

\begin{figure*}
\includegraphics[width=\textwidth]{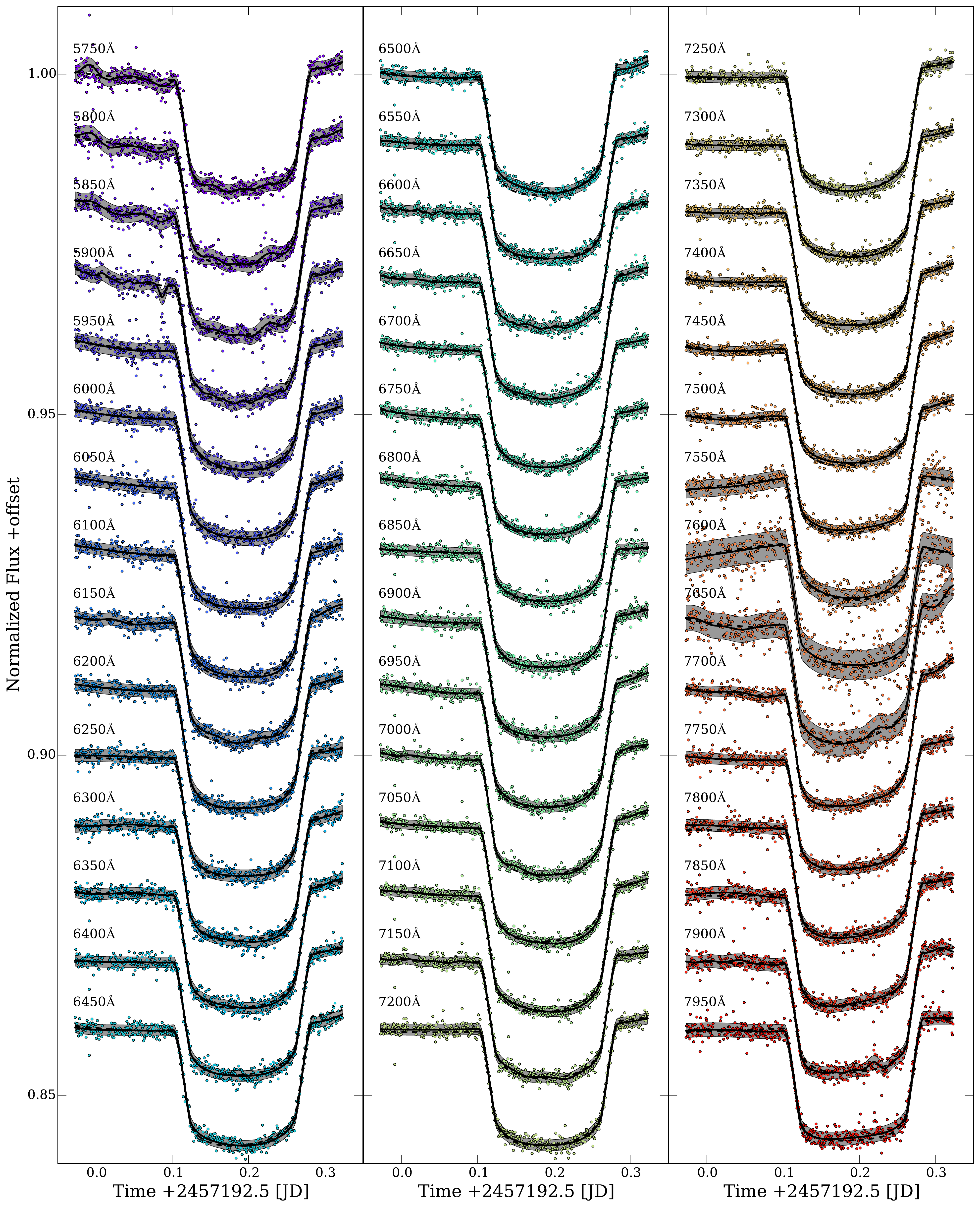}
\caption{Spectrophotometric light curves for the transit of WASP-17b, produced with bins of 100\AA~width after the common mode correction (CMC), shifted for clarity. The solid black line for each plot is the best fit transit systematics model and the grey areas are the 1-$\sigma$ representations of those solutions. The dashed lines represent transit models derived with the optimised parameters of each light curve. The central wavelength of each channel used to produce the light curve is given next to each plot.  The light curves excluded from transmission spectroscopy have been included here for completeness.}
\label{fig:100 spectro_LC}
\end{figure*}

\subsection{Transit Light Curve Fitting} \label{sec:GP model}

Once the final normalised broadband and spectrophotometric light curves are produced (see Figs. \ref{fig:raw LCs}, \ref{fig:broadband_LC} and \ref{fig:100 spectro_LC}), we first infer all the transit parameters, as well as the noise characteristics of the broadband light curve.  Subsequently, the same procedure is repeated for the spectrophotometric light curves with only the wavelength dependent
transit parameters and noise components inferred, while others are fixed to the white light curve solution, as it represents the light curve with the highest SNR, and hence the most accurate solutions.

We fit all the light curves using a Gaussian Process (GP) model, whose application to modelling systematics in time-series data has been introduced by \cite{Gibson2012b}. For determination of the GP model we have made use of their {\tt GeePea} module, as well as the {\tt Infer} module to implement Bayesian inference routines\footnote{The modules used are available at the following repository: \\ \url{https://github.com/nealegibson}}.

\begin{figure}
\includegraphics[width=\linewidth,height=9cm]{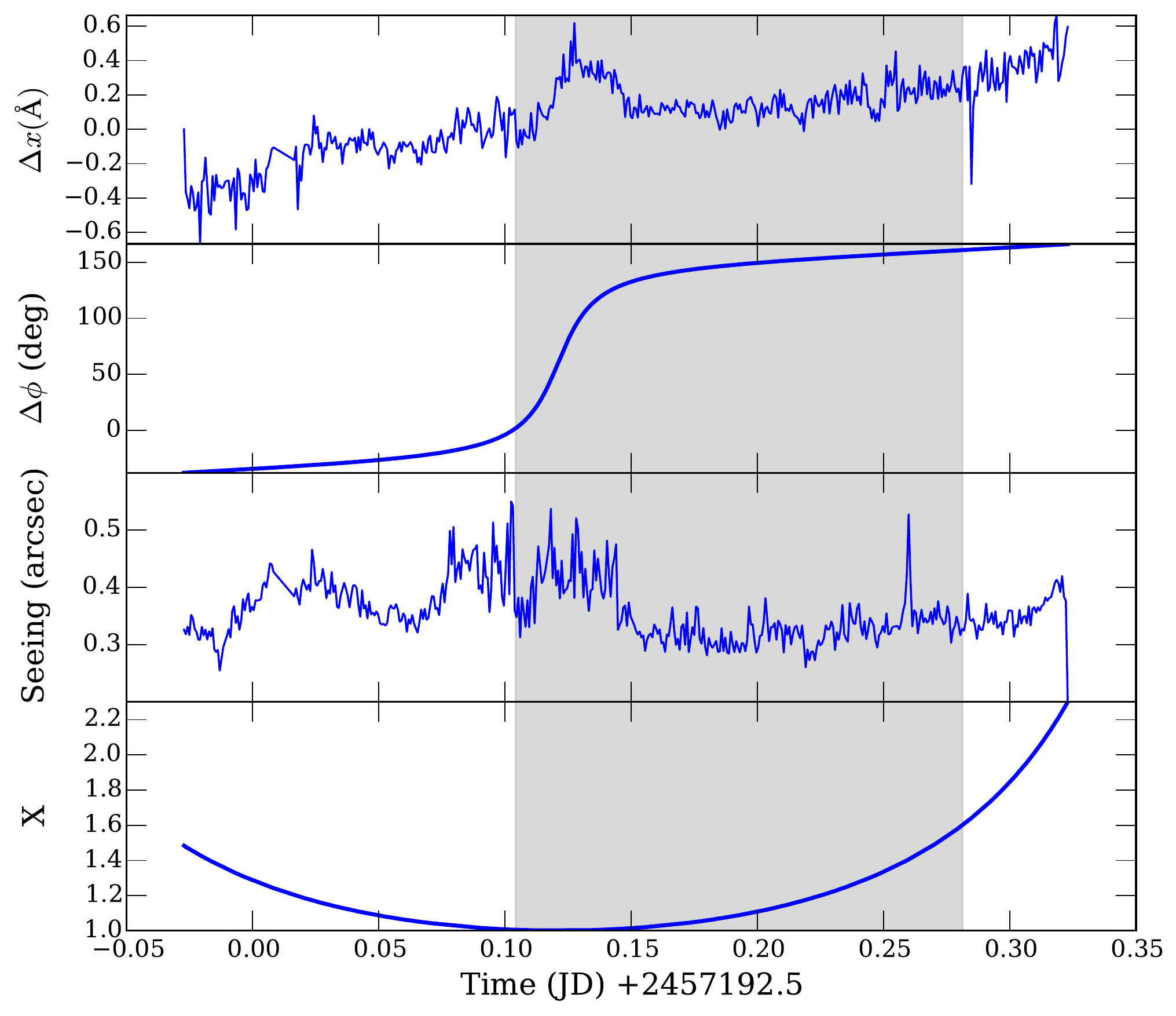}
\caption{Some of the optical state parameters extracted from the data. From the top, $\Delta x$ is the spectral drift relative to the first exposure, calculated using cross-correlation of the raw spectra, $\Delta \phi$ is the instrument rotation angle, Seeing is measured as the product of FWHM from the Gaussian profile along the spatial axis and the pixel scale in arcsec and $X$ is the airmass. These parameters are extracted to check for any possible correlation with the systematics in the light curve.  The grey--shaded region shows the duration of transit.}
\label{fig:optical state}
\end{figure}

A GP is essentially a non-parametric approach to regression analysis that does not necessarily make any assumptions about the data, nor does it impose a deterministic model on the data\footnote{This fact is essential in capturing the true impact of instrumental systematics in inferring the transit parameters.}. It is simply a collection of random variables, any finite subset of which has a joint Gaussian distribution \citep{Gibson2012b}. In this framework, each light curve is modelled as a GP of the analytical transit function, following the formalism of \cite{Mandel2002}, and the covariance matrix representing the noise characteristics of the data:
\begin{equation}\label{eq:GP}
f(t,\boldsymbol{\phi},\boldsymbol{\theta}) \sim \mathcal{GP}\left[T(t,\boldsymbol{\phi}),\boldsymbol{\Sigma}(t,\boldsymbol{\theta})\right]
\end{equation}
\noindent with $f$ being the flux measurements and $t$ the time of each observation. Here $T$ is the analytical transit model that is a function of $t$ and the transit parameters $\boldsymbol{\phi}$. In addition, $\boldsymbol{\Sigma}$ represents the covariance matrix, whose kernel has time as its only input, with parameters $\boldsymbol{\theta}$ also known as the \textit{hyperparameters}.  We also extracted as many physical variants, also known as the \textit{optical state parameters}, as available to search for any correlations between them and the residuals of a transit model fit to the data.  However, due to lack of any correlation none of this information was used in describing the GP model.  Some of such extracted information are shown in Fig. \ref{fig:optical state}.

The covariance matrix in Eq. \ref{eq:GP} is given by:
\begin{equation}
\Sigma_{ij} = k(t_i,t_j,\boldsymbol{\theta}) + \delta_{ij}\sigma^2_w = \zeta^2 \exp \left(-\frac{(t_i-t_j)^2}{2l^2}\right) + \delta_{ij}\sigma^2_w
\end{equation}

\noindent where $\zeta$ is the output scale describing the GP's variance, $l$ is the length scale that determines how smooth the function is, $\delta$ is the Kronecker delta ensuring the addition of white noise to the diagonal of the covariance matrix and $\sigma_w$ is that Poisson noise.  For the covariance function or the kernel, $k(t_i,t_j,\boldsymbol{\theta})$, we have chosen the \textit{Squared Exponential (SE)} kernel \citep[e.g.][]{Wilson2013} due to smoothness of the function over the parameters.  It must be emphasised that although selecting a kernel is analogous to choosing a predetermined parametric model, it allows for a much more flexible treatment of noise than any parametric form, as well as being intrinsically Bayesian.

We also define priors for all the hyperparameters of the noise model, ensuring that their values are positive, and together with the mean transit function parameters, optimise them with a Nelder-Mead simplex algorithm \citep{Glaudell1965}. To obtain the posterior joint probability distributions, these are multiplied by the marginal likelihoods to produce the joint posterior probability distribution.  In other words:
\begin{equation}
\ln P(\boldsymbol{\phi},\boldsymbol{\theta}|\boldsymbol{x}) = \ln P(\boldsymbol{\phi},\boldsymbol{\theta}) + \ln P(\boldsymbol{x}|\boldsymbol{\phi},\boldsymbol{\theta})
\end{equation}

We then use Monte Carlo Markov Chains (MCMC) to explore this posterior distribution and produce the final marginal probabilities for each of the parameters (for both the transit and the noise models). Essentially, we just assume uniform, uninformative priors for all the parameters with few restrictions, those being that the impact parameter\footnote{This is due to the degeneracy with the system scale.} and GP hyperparameters are positive and the sum of the two limb darkening coefficients of the quadratic law are less than unity to ensure positive brightness across the entire stellar surface. For all the light curves, we ran 4 MCMC chains with 100 walkers of 100000 iterations.  Once the chains are computed, we extract the marginalised posteriors for the free parameters to check convergence and any correlations. Examples of such correlation plots and posterior probability distributions are shown in Fig. \ref{fig:correlations} for the broadband and one spectrophotometric light curve.

\section{Results}
\subsection{Broadband Light Curve Models}

We first model the broadband or white light curve, which is the integral of fluxes for the largest possible common wavelength domain of the target and the comparison, namely from 5700 to 8000\AA, shown in Fig. \ref{fig:broadband_LC} modelled with the quadratic limb darkening law \citep{Kopal1950}.  In order to ensure that the choice of limb darkening law does not have a significant impact upon the inference of transit parameters, we initially model the white light curve using both the quadratic $I_{quad}(\mu)$ and the non-linear law $I_{nl}(\mu)$ \citep{Claret2000}, also known as the four-parameter law, given by:
\begin{eqnarray}
\frac{I_{quad}(\mu)}{I(1)} &=& 1 - \sum_{i=1}^2 c_i(1-\mu)^i \\
\frac{I_{nl}(\mu)}{I(1)} &=& 1 - \sum_{i=1}^4 l_i(1-\mu^{i/2})
\end{eqnarray}
\noindent respectively, where $\mu$ (=$\cos \theta$) is the cosine of the angle between the observer and the normal to the stellar surface.  These formulations have as particular cases, the linear law \citep[][$c_2$=$0$ in the quadratic law]{Schwarzschild1906} and the three-parameter law \citep[][$l_1$=$0$ in the non-linear law]{Sing2009}. In order to initially examine the impact of the chosen limb darkening law on the inferred planetary radius, we model the broadband light curve with the aforementioned laws, taking a variety of approaches for coefficient determination.  Beyond the quadratic law there is no improvement made to the fit when using a more complex law and there are no significant changes in the inferred transit parameters.  Furthermore, there exist high levels of correlation between the fitted parameters of those laws.  This together with the added unjustified complexity means that the quadratic law is chosen to model the radial intensity variations across the stellar disk.

In our modelling process both coefficients of the quadratic limb darkening law are fitted for in the white and spectral channels, as the stellar intensity profile is expected to be colour--dependent.  In determination of these coefficients we impose some informative priors.  The first of those restrictions is that the sum of their values is less than unity to ensure a positive value for the stellar surface intensity across the entire disk.  Initially, we calculate coefficient values for all channels using the {\tt PyLDTk}\footnote{Available from \url{https://github.com/hpparvi/ldtk}} package \citep{Parviainen2015} that uses the stellar atmosphere tables of \cite{Husser2013} generated with the {\tt PHOENIX} code.  We initially use these values to construct Gaussian priors for the inference of the coefficient values from the MCMC analysis, using the calculated values as mean and twice the error as the standard deviation.  However, this approach is rather restrictive and leads to systematic errors in the final fit.  Hence, we use these calculated values to construct a global and less informative prior given by:
\begin{equation*}
\ln P(c_1,c_2) = \begin{cases}
	-\infty & \text{if $\left(c_1 + c_2 > 1\right)$},\\
	-\infty & \text{if $\left(c_1<0 \vee c_1>1\right)$},\\
	-\infty & \text{if $\left(c_2<-1 \vee c_2>1\right)$},\\
	0 & \text{otherwise}.
	\end{cases}
\end{equation*}

The optimised parameters and the noise characteristics from fitting the broadband light curve, with the aboved mentioned setup, are given in Table \ref{tab:broadband fit}. Through the combination of multiple reference stars, the colour--dependent slope in the white light curve dispears, and hence we do not include a polynomial fit of the $\boldsymbol{\mathrm{f}}_{\mathrm{oot}}$ points.  This, however, is not the case for the narrowband channels.  These inferred planetary transit parameters from the MCMC analysis are consistent with those reported previously \citep{Anderson2010,Anderson2011,Southworth2012,Bento2014}.

\begin{table}
\centering
\caption{Parameters from the MCMC analysis of the broadband light curve, given for the fully marginalised Gaussian Process noise model.}
\begin{tabular}{l l}
\hline \hline
Parameter (inferred)& Value\\
\hline
Mid-transit, T$_{\mathrm{c}}$ (JD) +2400000 & $57192.69266\pm0.00028$\\
T$_{\mathrm{c}}$, Heliocentric corr. (HJD) & $57192.69798\pm0.00028$\\
Period, P (days) & $3.735438^a$ (fixed)\\
Scaled semi-major axis, $a/R_{\star}$ &$7.025\pm0.146$\\
Relative planetary radius, $R_p/R_{\star}$ & $0.12345\pm0.00109$\\
Impact parameter, $b$ & $0.361\pm0.069$ \\
Linear LD coefficient, $c_1$ & $0.167 \pm 0.136$\\
Quadratic LD coefficient, $c_2$ & $0.356\pm0.249$\\
$f_{oot}$ & $1.0$ (fixed) \\
$T_{grad}$ & $0.0$ (fixed)\\
GP max. covariance, $\zeta$ (ppm) & $545\pm259$\\
Length scale, $l$ & $0.019\pm0.009$\\
White noise, $\sigma_w$ (ppm)& $545\pm19$\\
Eccentricity, $e$ & 0.0 (fixed)\\
Argument of periapsis, $\omega~(^{\circ})$ & 90.0 (fixed) \\
\hline
Parameter (derived)& Value\\
\hline
Semi-major axis, $a$ (AU) & $0.0513\pm0.00290$\\
Inclination, $i$ & $87.06^{\circ~+0.61}_{~~~-0.63}$\\
$M_{\star}$ $(M_{\odot})$& $1.306\pm0.026^a$\\
$R_{\star}$ $(R_{\odot})$& $1.572\pm0.056^a$\\
$M_{\mathrm{p}}$ $(M_{\mathrm{J}})$& $0.486\pm0.032^a$\\
$R_{\mathrm{p}}$ $(R_{\mathrm{J}})$& $1.747\pm0.078$\\
$\rho_{\mathrm{p}}$ $(\mathrm{g~cm^{-3}})$& $0.121\pm0.024$ $(0.091~\rho_{\mathrm{J}})$\\
$g_{\mathrm{p}}$ $(\mathrm{m~s^{-2}})$ &  $3.948\pm0.612$ \\
\hline
\multicolumn{2}{l}{$^{a}$~{\scriptsize Value from \cite{Anderson2011}.}}
\end{tabular}
\label{tab:broadband fit}
\end{table}

\subsection{Transmission Spectrum} \label{sec:trans spec}

\begin{figure*}
\includegraphics[width=\textwidth]{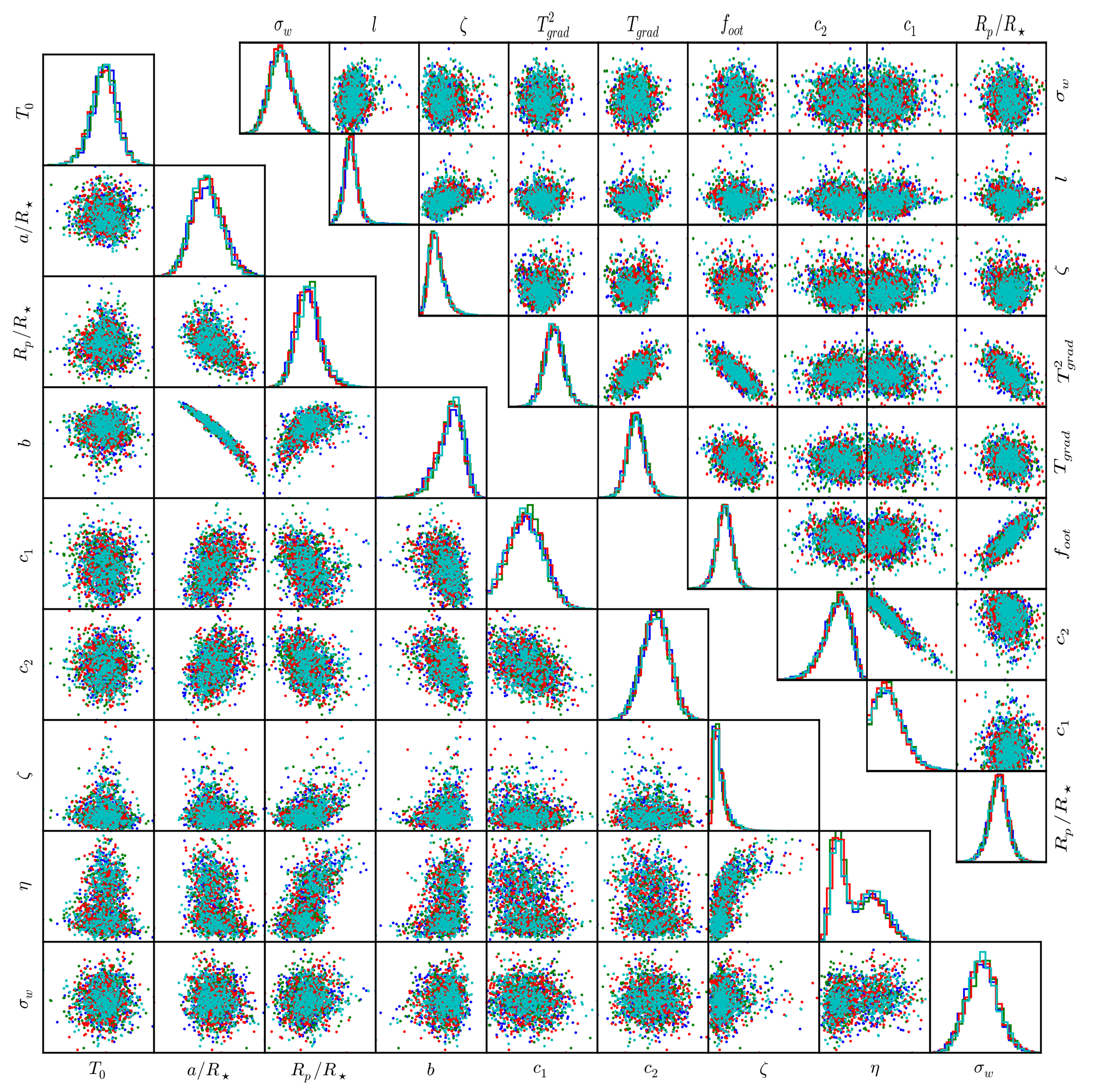}
\caption{Parameter correlations for the broadband (bottom left) and a spectrophotometric (top right) light curve from the MCMC analysis, for the quadratic limb darkening law. The well-documented degeneracy between the impact parameter, $b$, and the scaled semi-major axis, $a/R_{\star}$ is also quite evident. The derived posterior probability distribution for each parameter is also given at the end of each column. The four different colours represent samples from the independent MCMC chains.}
\label{fig:correlations}
\end{figure*}

Once the system parameters are determined from the analysis of the broadband light curve, we model the spectrophotometric light curves in the same way but with the non-wavelength dependent parameters fixed to the best solution from the white light curve, given in Table \ref{tab:broadband fit}. Since this light curve represents the time-series flux measurements with the highest SNR, its solutions are those most trusted. In modelling the spectroscopic light curves we fit for the scaled planetary radius, $R_p/R_{\star}$, both coefficients of the limb darkening law, $c_1$ and $c_2$, as well as the three quantities describing the noise statistics of each data set, $\zeta$, $l$ and $\sigma_w$, explained previously.  In addition to those, in the modelling of spectrophotometric light curves, we also simultaneously fit for three coefficients of a second order polynomial which approximates long term trends in the differential flux.  Such variations are due to larger colour differences between the target and the comparison in smaller bins as compared to the broadband bin, where such effects are not apparent.  This effect is a function of airmass and spectral type described well with a polynomial of second order, which can be seen in some of the light curves in Fig. \ref{fig:100 spectro_LC}. An example of the correlation plots for the free parameters of one spectral light curve is shown in the upper panel of Fig. \ref{fig:correlations}. 

In addition to this approach, we also correct the spectrophometric light curves for a common mode of the systematic noise that is shared across all the spectral channels.  This is done by dividing all the spectrophotometric light curves by the residuals of best fit transit model to the broadband light curve, shown as the blue data points at the bottom of Fig. \ref{fig:broadband_LC}.  This residual data is simply obtained by the division of the transit data by the analytical model (blue solid line in the top figure of the same plot).  Subsequently, we model these light curves again with the same procedure, where the red noise is modelled as a GP due to any residual systematic noise after this \textit{common mode correction} (referred to as CMC hereafter).  These light curves together with the transit and GP noise models are shown in Fig. \ref{fig:100 spectro_LC}.

\begin{figure*}
\includegraphics[width=\textwidth]{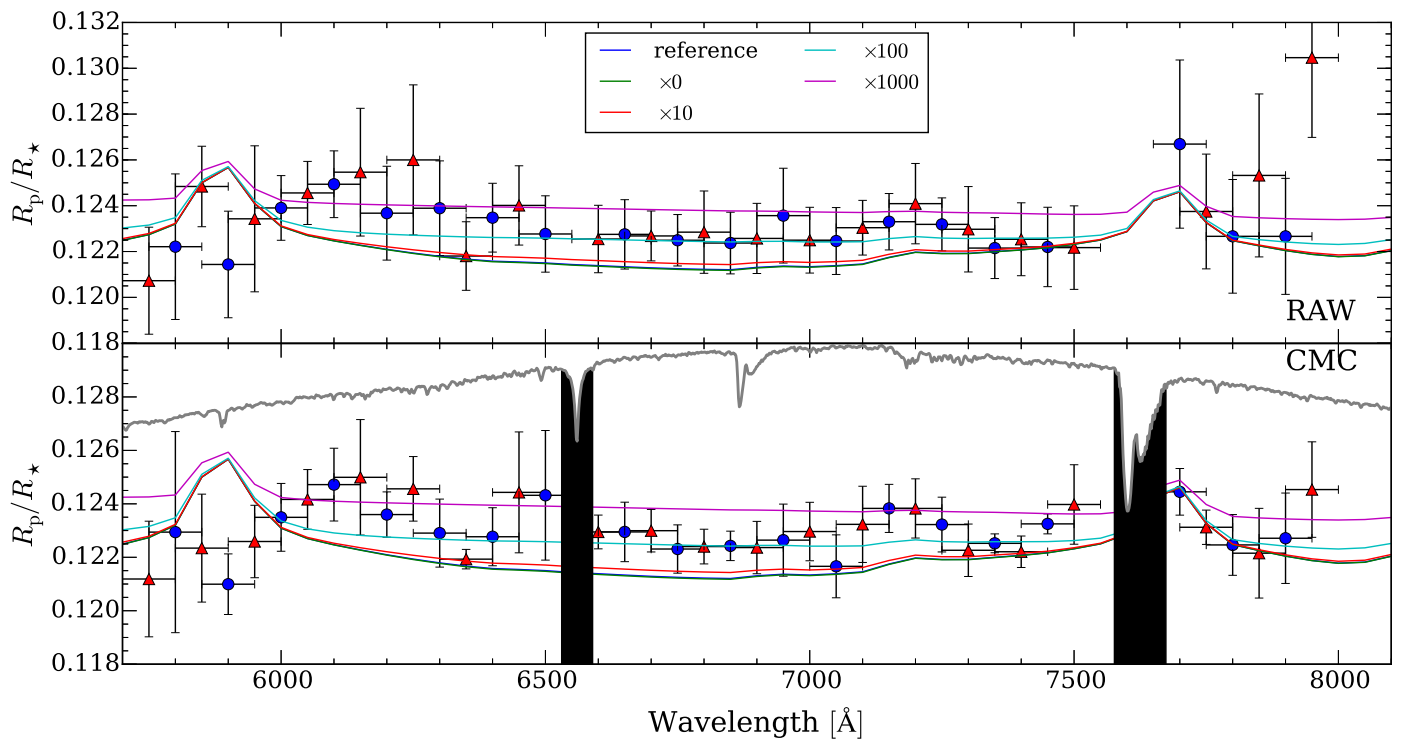}
\caption{Transmission spectrum of WASP-17b produced from both without (\textit{top}) and with (\textit{bottom}) the \textit{Common Mode Correction} of the spectrophotometric light curves.  Blue circles and red triangles are results from light curves made with 100\AA~bins, plotted as two separate sets to show transmission spectra with unique data points.  Furthermore, synthetic transmission models for the reference aerosol profile scaled by factors of 0, 10, 100 and a 1000 are also over plotted. These models are binned within 100\AA~channels to match the data.  In the CMC plot, an example of a WASP-17 spectrum is over plotted in grey.  Two regions of telluric absorption have been shaded black, where we avoid reporting results due to the low SNR.  It must be stressed that the models are not fitted to the data and simply just over plotted.  Hence, we do not try the retrieval approach since the wavelength coverage of the spectrum is too limited for this approach.}
\label{fig:trans spec}
\end{figure*}

The variations of relative planetary radii as a function of channel central wavelength, $\lambda_c$, are derived from this modelling process, which represents the transmission spectrum of WASP-17b. A set of results from modelling of both data sets (initial and CMC), including relative radii and limb darkening coefficients, are tabulated in table \ref{tab:trans results}.  The transmission spectrum is shown in Fig. \ref{fig:trans spec} for the two employed procedures.

\begin{figure*}
\includegraphics[width=\textwidth]{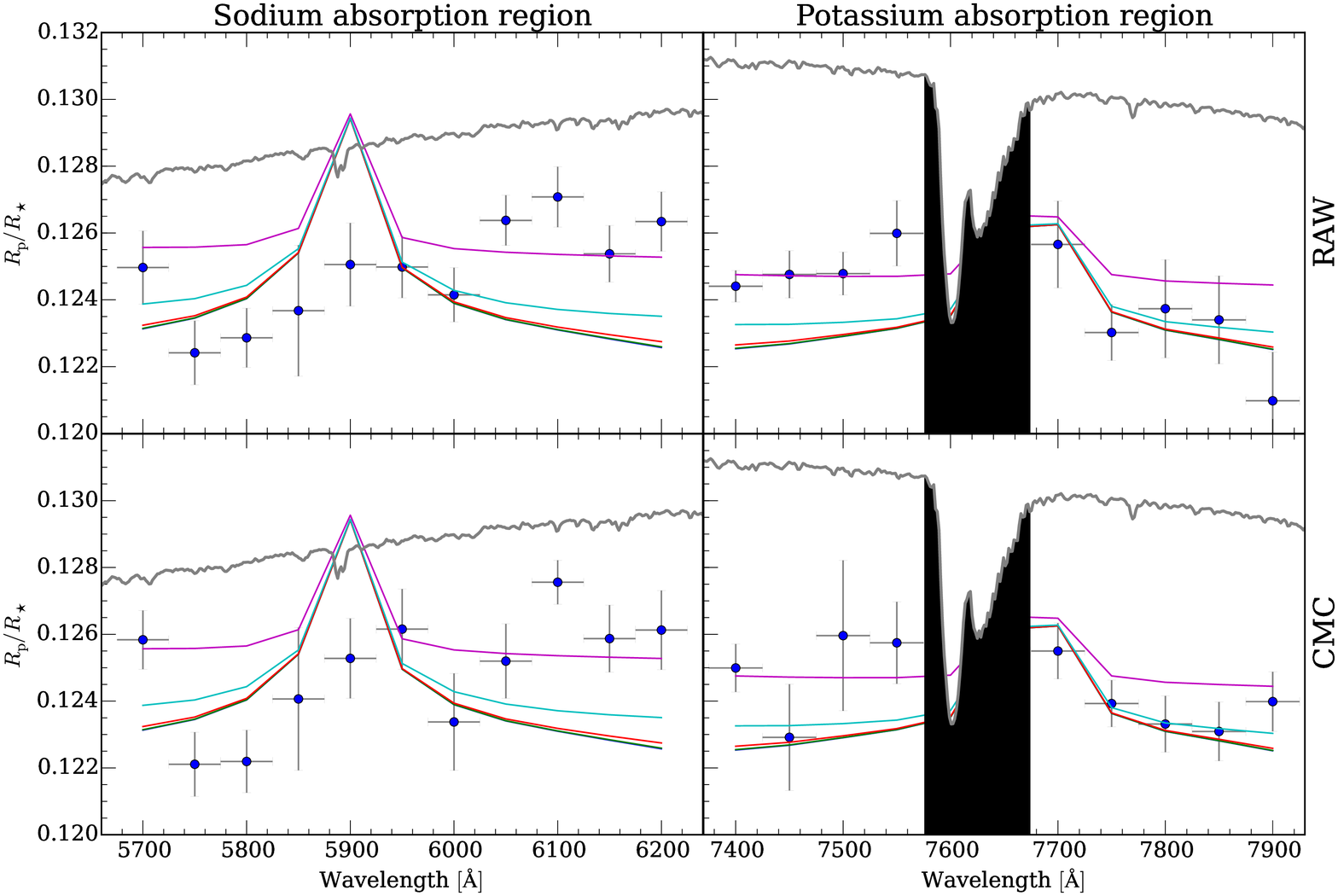}
\caption{Transmission spectrum reproduced with 50\AA--wide bins, as compared to Fig. \ref{fig:trans spec}, for domains where possible presence of optical absorbers, such as sodium (\textit{left column}) and potassium (\textit{right column}), would result in an increased planetary radius. The theoretical atmospheric models are also over-plotted in addition to the inferred radii and their colours correspond to those in Fig. \ref{fig:trans spec}, which have been again binned to the resolution of the spectrum and plotted for 1.6 scale heights, as was done for the overall spectrum. It must be noted that these models have not been fitted to the data points and the transmission spectrum data have not been used as \textit{a priori} for the production of the model atmospheres.  The black shaded region in the right column highlights the region where telluric absorption due to the $O_2$ (A) lines introduces significant systematics in the light curves produced in this region.}
\label{fig:trans_spec_absorption}
\end{figure*}

One possible feature of these spectra is the indication of enhanced absorption in the exoplanetary atmosphere in the range of 7650--7800\AA, which is possibly due to a combination of the telluric $O_2$(A) lines and the Potassium doublet (7665\AA~\& 7699\AA) in the exoplanet's atmosphere.  This point will be discussed further in §\ref{sec:atm model}.  To look for possible optical absorbers in the atmosphere of WASP-17b, we also produce narrow band light curves with 50\AA~integration bins around Sodium and Potassium absorption regions, which are shown in Fig. \ref{fig:trans_spec_absorption}.

\begin{table*}
\centering
\caption{Results from modelling spectrophotometric light curves produced with 100\AA~integration bins, for both data sets.}
\begin{tabular}{l c c c | c c c}
\hline \hline
\multicolumn{1}{c}{Band}&\multicolumn{6}{c}{Spectrophotometric LC set}\\ \cline{2-7}
\multicolumn{1}{c}{Centre} & \multicolumn{3}{c}{RAW} & \multicolumn{3}{c}{CMC} \\ \cline{2-7}
\multicolumn{1}{c}{[\AA]}& $R_p/R_{\star}$ & $c_1$ & $c_2$ & $R_p/R_{\star}$ & $c_1$ & $c_2$ \\
\hline
5750	 & 	0.1207$\pm$0.0023	 & 	0.0866$\pm$0.3687	 & 	0.5760$\pm$0.6236	 & 	0.1212$\pm$0.0022	 & 	0.0921$\pm$0.0604	 & 	0.5687$\pm$0.1396\\ 
5800	 & 	0.1222$\pm$0.0032	 & 	0.0902$\pm$0.1368	 & 	0.5799$\pm$0.2794	 & 	0.1229$\pm$0.0038	 & 	0.0953$\pm$0.0630	 & 	0.5805$\pm$0.1485\\ 
5850	 & 	0.1248$\pm$0.0018	 & 	0.0880$\pm$0.1404	 & 	0.6034$\pm$0.2849	 & 	0.1223$\pm$0.0020	 & 	0.0833$\pm$0.0605	 & 	0.5959$\pm$0.1419\\ 
5900	 & 	0.1214$\pm$0.0023	 & 	0.1415$\pm$0.1136	 & 	0.5714$\pm$0.2347	 & 	0.1210$\pm$0.0011	 & 	0.1436$\pm$0.0753	 & 	0.5685$\pm$0.1666\\ 
5950	 & 	0.1234$\pm$0.0032	 & 	0.2407$\pm$0.1366	 & 	0.4574$\pm$0.2537	 & 	0.1226$\pm$0.0014	 & 	0.2300$\pm$0.0855	 & 	0.4552$\pm$0.1636\\ 
6000	 & 	0.1239$\pm$0.0014	 & 	0.1706$\pm$0.0846	 & 	0.4926$\pm$0.1748	 & 	0.1235$\pm$0.0013	 & 	0.1679$\pm$0.0722	 & 	0.4825$\pm$0.1373\\ 
6050	 & 	0.1246$\pm$0.0014	 & 	0.1195$\pm$0.0665	 & 	0.5695$\pm$0.1387	 & 	0.1242$\pm$0.0011	 & 	0.1173$\pm$0.0613	 & 	0.5533$\pm$0.1213\\ 
6100	 & 	0.1249$\pm$0.0015	 & 	0.1511$\pm$0.0833	 & 	0.4923$\pm$0.1803	 & 	0.1247$\pm$0.0014	 & 	0.1529$\pm$0.0760	 & 	0.4878$\pm$0.1544\\ 
6150	 & 	0.1255$\pm$0.0028	 & 	0.1618$\pm$0.1428	 & 	0.5370$\pm$0.2924	 & 	0.1250$\pm$0.0022	 & 	0.1539$\pm$0.0836	 & 	0.5249$\pm$0.1644\\ 
6200	 & 	0.1237$\pm$0.0020	 & 	0.1195$\pm$0.0590	 & 	0.6025$\pm$0.1184	 & 	0.1236$\pm$0.0008	 & 	0.1018$\pm$0.0557	 & 	0.5837$\pm$0.1114\\ 
6250	 & 	0.1260$\pm$0.0033	 & 	0.1030$\pm$0.1534	 & 	0.6200$\pm$0.2926	 & 	0.1246$\pm$0.0012	 & 	0.0736$\pm$0.0472	 & 	0.5971$\pm$0.1029\\ 
6300	 & 	0.1239$\pm$0.0021	 & 	0.0944$\pm$0.0639	 & 	0.6260$\pm$0.1266	 & 	0.1229$\pm$0.0013	 & 	0.0784$\pm$0.0480	 & 	0.6074$\pm$0.1017\\ 
6350	 & 	0.1218$\pm$0.0015	 & 	0.1252$\pm$0.0657	 & 	0.5344$\pm$0.1347	 & 	0.1219$\pm$0.0004	 & 	0.1314$\pm$0.0625	 & 	0.5205$\pm$0.1243\\ 
6400	 & 	0.1235$\pm$0.0015	 & 	0.1463$\pm$0.0663	 & 	0.4382$\pm$0.1358	 & 	0.1228$\pm$0.0011	 & 	0.1424$\pm$0.0616	 & 	0.4168$\pm$0.1272\\ 
6450	 & 	0.1240$\pm$0.0017	 & 	0.1813$\pm$0.0897	 & 	0.3792$\pm$0.1746	 & 	0.1244$\pm$0.0023	 & 	0.1767$\pm$0.0821	 & 	0.3679$\pm$0.1587\\ 
6500	 & 	0.1228$\pm$0.0017	 & 	0.1932$\pm$0.0940	 & 	0.3895$\pm$0.1896	 & 	0.1243$\pm$0.0024	 & 	0.1874$\pm$0.0893	 & 	0.3776$\pm$0.1579\\ 
6600	 & 	0.1225$\pm$0.0015	 & 	0.1168$\pm$0.0552	 & 	0.4415$\pm$0.1206	 & 	0.1229$\pm$0.0006	 & 	0.1067$\pm$0.0508	 & 	0.4298$\pm$0.1094\\ 
6650	 & 	0.1227$\pm$0.0015	 & 	0.1600$\pm$0.0593	 & 	0.4636$\pm$0.1274	 & 	0.1229$\pm$0.0011	 & 	0.1589$\pm$0.0703	 & 	0.4542$\pm$0.1478\\ 
6700	 & 	0.1227$\pm$0.0011	 & 	0.2447$\pm$0.0866	 & 	0.3224$\pm$0.1793	 & 	0.1230$\pm$0.0008	 & 	0.2325$\pm$0.0645	 & 	0.3166$\pm$0.1363\\ 
6750	 & 	0.1225$\pm$0.0011	 & 	0.1929$\pm$0.0707	 & 	0.3577$\pm$0.1430	 & 	0.1223$\pm$0.0009	 & 	0.1913$\pm$0.0609	 & 	0.3463$\pm$0.1212\\ 
6800	 & 	0.1228$\pm$0.0018	 & 	0.1136$\pm$0.1253	 & 	0.4925$\pm$0.2640	 & 	0.1224$\pm$0.0006	 & 	0.1083$\pm$0.0488	 & 	0.4858$\pm$0.1047\\ 
6850	 & 	0.1224$\pm$0.0013	 & 	0.1323$\pm$0.1089	 & 	0.4441$\pm$0.2315	 & 	0.1224$\pm$0.0006	 & 	0.1284$\pm$0.0490	 & 	0.4453$\pm$0.1030\\ 
6900	 & 	0.1226$\pm$0.0015	 & 	0.1553$\pm$0.0876	 & 	0.4365$\pm$0.1818	 & 	0.1224$\pm$0.0010	 & 	0.1530$\pm$0.0624	 & 	0.4320$\pm$0.1255\\ 
6950	 & 	0.1236$\pm$0.0021	 & 	0.1792$\pm$0.0937	 & 	0.4279$\pm$0.1759	 & 	0.1226$\pm$0.0013	 & 	0.1730$\pm$0.0728	 & 	0.4104$\pm$0.1383\\ 
7000	 & 	0.1225$\pm$0.0014	 & 	0.2107$\pm$0.0926	 & 	0.3906$\pm$0.1845	 & 	0.1230$\pm$0.0011	 & 	0.2049$\pm$0.0846	 & 	0.3826$\pm$0.1751\\ 
7050	 & 	0.1225$\pm$0.0015	 & 	0.1615$\pm$0.0776	 & 	0.5082$\pm$0.1579	 & 	0.1217$\pm$0.0012	 & 	0.1498$\pm$0.0638	 & 	0.4881$\pm$0.1244\\ 
7100	 & 	0.1230$\pm$0.0012	 & 	0.1782$\pm$0.0843	 & 	0.4435$\pm$0.1782	 & 	0.1232$\pm$0.0014	 & 	0.1830$\pm$0.0839	 & 	0.4398$\pm$0.1704\\ 
7150	 & 	0.1233$\pm$0.0012	 & 	0.2003$\pm$0.0837	 & 	0.3785$\pm$0.1791	 & 	0.1238$\pm$0.0009	 & 	0.1963$\pm$0.0735	 & 	0.3733$\pm$0.1527\\ 
7200	 & 	0.1241$\pm$0.0018	 & 	0.1881$\pm$0.0879	 & 	0.3795$\pm$0.1697	 & 	0.1238$\pm$0.0011	 & 	0.1774$\pm$0.0674	 & 	0.3754$\pm$0.1308\\ 
7250	 & 	0.1232$\pm$0.0012	 & 	0.2840$\pm$0.0928	 & 	0.1941$\pm$0.1906	 & 	0.1232$\pm$0.0010	 & 	0.2777$\pm$0.0662	 & 	0.1884$\pm$0.1334\\ 
7300	 & 	0.1230$\pm$0.0019	 & 	0.2174$\pm$0.0878	 & 	0.3023$\pm$0.1639	 & 	0.1223$\pm$0.0010	 & 	0.2045$\pm$0.0626	 & 	0.2919$\pm$0.1247\\ 
7350	 & 	0.1222$\pm$0.0013	 & 	0.1704$\pm$0.0809	 & 	0.4045$\pm$0.1555	 & 	0.1225$\pm$0.0004	 & 	0.1567$\pm$0.0549	 & 	0.3862$\pm$0.1014\\ 
7400	 & 	0.1225$\pm$0.0016	 & 	0.1314$\pm$0.0922	 & 	0.4483$\pm$0.1680	 & 	0.1222$\pm$0.0006	 & 	0.1132$\pm$0.0468	 & 	0.4338$\pm$0.0899\\ 
7450	 & 	0.1222$\pm$0.0017	 & 	0.2164$\pm$0.0996	 & 	0.3218$\pm$0.1787	 & 	0.1233$\pm$0.0004	 & 	0.1995$\pm$0.0640	 & 	0.3082$\pm$0.1105\\ 
7500	 & 	0.1222$\pm$0.0018	 & 	0.1780$\pm$0.0511	 & 	0.3904$\pm$0.1049	 & 	0.1240$\pm$0.0015	 & 	0.1652$\pm$0.0797	 & 	0.3758$\pm$0.1461\\ 
7700	 & 	0.1267$\pm$0.0037	 & 	0.1024$\pm$0.1140	 & 	0.4673$\pm$0.1980	 & 	0.1244$\pm$0.0009	 & 	0.1015$\pm$0.0831	 & 	0.4703$\pm$0.1635\\ 
7750	 & 	0.1237$\pm$0.0025	 & 	0.0691$\pm$0.0975	 & 	0.5359$\pm$0.1698	 & 	0.1231$\pm$0.0006	 & 	0.0680$\pm$0.0396	 & 	0.5373$\pm$0.0822\\ 
7800	 & 	0.1227$\pm$0.0025	 & 	0.1053$\pm$0.0932	 & 	0.4935$\pm$0.1570	 & 	0.1225$\pm$0.0011	 & 	0.1004$\pm$0.0621	 & 	0.4931$\pm$0.1110\\ 
7850	 & 	0.1253$\pm$0.0036	 & 	0.1225$\pm$0.1267	 & 	0.4833$\pm$0.2386	 & 	0.1222$\pm$0.0017	 & 	0.1229$\pm$0.0765	 & 	0.4814$\pm$0.1331\\ 
7900	 & 	0.1227$\pm$0.0025	 & 	0.1632$\pm$0.0769	 & 	0.3366$\pm$0.1724	 & 	0.1227$\pm$0.0017	 & 	0.1591$\pm$0.0918	 & 	0.3217$\pm$0.1978\\ 
7950	 & 	0.1305$\pm$0.0035	 & 	0.1006$\pm$0.1259	 & 	0.3853$\pm$0.2108	 & 	0.1245$\pm$0.0018	 & 	0.0832$\pm$0.0587	 & 	0.3697$\pm$0.1072\\ 
\hline
\multicolumn{7}{l}{{\scriptsize \textit{Note} The corresponding CMC light curves are given in Figs. \ref{fig:100 spectro_LC}. The three transit parameters, taken as free during the MCMC analysis, namely the relative planetary}}\\ 
\multicolumn{7}{l}{~~~~~~{\scriptsize radius ($R_p/R_{\star}$) and the two coefficients of the quadratic limb darkening law ($c_1$ \& $c_2$) are shown. Other free components for each channel were the 3 coefficients}}\\
\multicolumn{7}{l}{~~~~~~{\scriptsize of the polynomial fit, as well as the 3 noise model parameters.}}
\end{tabular}
\label{tab:trans results}
\end{table*}

\subsection{Atmospheric Models}\label{sec:atm model}

We compare the derived transmission spectra of Fig. \ref{fig:trans spec} to especially adapted theoretical atmospheric spectral models also shown in the same figure.  These models are calculated with the assumption of isothermal profile using the equilibrium temperature of WASP-17b, as well as full heat redistribution and zero albedo. The only spectroscopically active gases present in the model atmosphere are H$_2$O, Na and K, with volume mixing ratios of $1 \times 10^{-3}$, $2.96 \times 10^{-6}$ and $2.4 \times 10^{-7}$, the latter 2 being solar values assuming H$_2$/He bulk abundances and the water mixing ratio being somewhat arbitrary. These mixing ratios are assumed independent of altitude.  We take the specific gravity to be non-constant and varying with altitude, as this probably would be significant for a puffy and low density planet like WASP-17b.  Furthermore, we assume that the aerosol density decays with altitude using a scale height similar to that of the gas.  To explore the impact of aerosol density, we have scaled the reference aerosol profile ($\times 1$), by factors of 0, 10, 100 and 1000, which are all shown in Fig. \ref{fig:trans spec} superimposed onto the derived transmission spectra.  In these cases the aerosol optical thickness of 1 at 0.8 micron is reached at pressure levels of $\sim$7, 1, 0.15 and 0.03 bars, respectively. This aerosol optical thickness is assumed to follow a Rayleigh-type law with a wavelength dependence of $\lambda^{-4}$. This law is valid when the size parameter $x$ (=$2\pi r_{\text{eff}}/\lambda$; where $r_{\text{eff}}$ is a measure of the particle radius) is moderately small and the imaginary part of the refractive index is also small. For instance, silicates, perovskite and silica have small imaginary parts for their refractive indices. This is a hard-coded assumption of the model, but is anyway consistent with the Rayleigh slope that is seen in numerous exoplanets. WASP-17b has a rather puffy composition, hence one could expect that small particles are easily kept aloft in the atmosphere.  Given the very unconstrained nature of the problem, we think that the treatment used here is judicious.  Finally, given the fact that the extinction coefficient decays exponentially with altitude, and the pressure levels at which maximal optical thickness is reached (i.e. $\tau$ =1), one reconstructs the aerosol extinction coefficient needed in the models\footnote{Note that the aerosol densities and cross sections are degenerate since only their product is required in transmission calculations.}. These theoretical atmospheric models are updated versions from \citet{Munoz2012}.

The plotted theoretical models have been shifted vertically to match the visible size of the planet, and scaled by $\sim$1.6 scale heights, a value which has been obtained through minimization of $\chi^2$ statistics using least squares, with transmission spectrum data points.

We calculate $\chi^2$ and significance statistics using the CMC spectrum of Fig. \ref{fig:trans spec} to establish whether one can discriminate between a clear or cloudy and opaque atmosphere.  To do this, the synthetic spectra are compared to a flat spectrum with a one scale height tolerance.  We calculate a reduced $\chi^2_r$ value of $\sim$1.19 for a flat spectrum, as compared to $\sim$1.07--1.11 for the various atmospheric models.  Hence, we rule out a flat spectrum at >$3\sigma$, with the $\Delta${\tt BIC} ($\gg$10)\footnote{Bayesian Information Criterion \citep{Schwarz1978}.} value providing further, \textit{very strong} evidence against it.  Details of the statistical analysis of the individual models are given in Table \ref{tab:model stats}.  From this analysis, we conclude that our transmission spectrum significantly rules out a flat spectrum and suggests an atmosphere clear of large cloud particles for WASP-17b, with possible presence of smaller scattering particles, as there is marginally better agreement of data with the models of increased aerosol density.
\begin{table}[t]
\centering
\caption{Calculated statistics for the various models over plotted with the data.  $A_H$ is the optimised number of scale heights.}
\begin{tabular}{| l | p{2.5cm}p{2.5cm}p{2.5cm} |}
\cline{2-4}
\multicolumn{1}{c}{} & \multicolumn{3}{|c|}{100\AA}\\ \hline 
Atmospheric model & \multicolumn{1}{c}{$\chi^2_r$} & \multicolumn{1}{c}{$\Delta${\tt BIC}} & \multicolumn{1}{c|}{$A_H$}\\
\hline
\textbf{Flat} & \multicolumn{1}{c}{\textbf{1.19}} & \multicolumn{1}{c}{--} & \multicolumn{1}{c|}{--}\\
Reference aerosol ($\times 0$) & \multicolumn{1}{c}{1.07} & \multicolumn{1}{c}{157} & \multicolumn{1}{c|}{0.47} \\
$\times 0$ & \multicolumn{1}{c}{1.07} & \multicolumn{1}{c}{157} & \multicolumn{1}{c|}{0.46} \\
$\times 10$ & \multicolumn{1}{c}{1.07} & \multicolumn{1}{c}{157} & \multicolumn{1}{c|}{0.52} \\
$\times 100$ & \multicolumn{1}{c}{1.07} & \multicolumn{1}{c}{157} & \multicolumn{1}{c|}{0.82} \\
$\times 1000$ & \multicolumn{1}{c}{1.11} & \multicolumn{1}{c}{155} & \multicolumn{1}{c|}{1.15} \\
\hline
\end{tabular}
\label{tab:model stats}
\end{table}

Following the recipe of \cite{Nascimbeni2013} and \cite{Mallonn2016}, we estimate the mean molecular weight of the atmosphere using a fitted Rayleigh slope, from the relation:
\begin{equation}
\text{H} = \frac{1}{\alpha} \frac{\text{d}R_p}{\text{dln}\lambda} = \frac{k_BT_{eq}}{\mu_mg_p}
\end{equation}
\noindent where H is the scale height of the atmosphere, $\alpha$=$-4$ for Rayleigh scattering, $k_B$ is the Boltzman's constant, $T_{eq}$ the exo-atmosphere's equilibrium temperature, $\mu_m$ is the mean molecular weight of that atmosphere and $g_p$ is the planetary surface gravity. We estimate the Rayleigh scattering slope using the CMC transmission spectrum.  Due to larger uncertainties at either extreme of the spectrum, as well as telluric effects and possible absorption by exoatmospheric gases, we only include data points from 6100 to 7500\AA~for the calculations. The slope is subsequently obtained through a weighted least squares fitting method, giving the best fit value of:
\begin{equation}
\frac{\text{d}R_p}{\text{dln}\lambda} = \left(-0.00664 \pm 0.00233\right) \times R_{\star}
\end{equation}
\noindent Using the stellar radius value from \cite{Anderson2011}, we obtain a mean molecular weight, $\mu_g$, of $2.05\pm0.79$ a.m.u. for Bond albedo of 0, and $1.88\pm0.73$ a.m.u. for $\alpha_B=0.3$, which has been shown to be a reasonable upper limit for hot gas giants \citep{Spiegel2010,Esteves2013}. These values are generally consistent with a mostly H$_2$ dominated atmosphere, but due to the large uncertainties in determination of Rayleigh scattering slope, they could also point to presence of H or He.  However, a lack of H-$\alpha$ absorption in the transmission spectra somewhat rules out the former.

We also perform a similar analysis for the additional spectra created around the sodium and potassium absorption lines with narrower bins of 50\AA, shown in Fig. \ref{fig:trans_spec_absorption} to determine the significance of any possible detection.  \cite{Wood2011} detected an increases in transit depth    of 1.46\%, 0.55\% and 0.49\% for 0.75\AA, 1.5\AA~and 3\AA~bin, respectively.  However, they report no significant detection for bins of 4\AA~and higher.  Hence our non-detection of increased absorption in the Na core with the 50\AA~bin is consistent with their results.  Similar conclusions were also made by \cite{Zhou2012}, where transit depth increase of 0.58\% was detected in the Na core using a 1.5\AA~bin.  Considering only the points around the sodium line in Fig. \ref{fig:trans_spec_absorption}, we calculate a marginal improvement ($\sim$1$\sigma$) to a fit, when a flat spectrum is compared to a model with sodium absorption included.  Presence of pressure broadening would point to a rather shallow temperature gradient, although more precise measurements of the planetary radius at higher resolution are required to confirm this result.  More recently, \cite{Sing2015} also detected increased absorption in the sodium core using the Space Telescope Imaging Spectrograph (STIS) instrument on--board the Hubble Space Telescope (HST), using a narrow bin.  Using very small integration bins placed on the core of the absorption line, probes the lower atmosphere \citep{Wood2011}, as opposed to larger bins where the diminished signal could be due to the presence of silicate and iron clouds increasing opacity \citep{Fortney2003}.  The absence of sodium from high altitudes can be explained for equilibrium temperatures below $\sim$1000 K, where atomic Na is lost to formation of various compounds and rained out of the upper atmosphere \citep{Burrows2000}, but this is not the case for WASP-17b.    \citet{Lavvas2014} provide yet further possible explanations for presence and/or absence of Na and K in giant exoplanet atmospheres.

Similarly we also search for potassium absorption in the atmosphere of WASP-17b, obtaining and analyzing light curves using 50\AA~integration bins.  Unfortunately, the majority of the potassium doublet core falls on top of the telluric $O_2$ (A) absorption lines, as shown in the right column of Fig. \ref{fig:trans_spec_absorption}, which introduce large systematics for light curves obtained in this region.  Hence, we avoid producing and analyzing transit light curves in this domain where the signal is diminished by earth's atmosphere.  However, since the potassium core falls close to the red edge of the telluric forbidden region, we are able to probe and detect the red wing of the pressure-broadened line.  Using the data points redwards of the telluric absorption in Fig. \ref{fig:trans_spec_absorption}, we find a 3-$\sigma$ improvement to the fit when a flat spectrum is compared to the atmospheric models including potassium.  To date only one previous work has looked at this domain of the spectrum \citep{Sing2015}, where the narrow bin placed at the core of the potassium line is inconclusive in determining its presence or absence.  Given that potassium is present in the atmosphere of WASP-17b, the pressure broadening of its wings would be consistent with what we observe for the sodium line.

\citet{Charbonneau2002} first reported detection of Na in HD209458b. Subsequent studies suggested that the Na absorption was weaker than estimated by theory \citep{Seager2000} and that the potassium signal was lacking \citep{Snellen2008} for this planet. Four factors have been suggested which could influence the spectral bands of the alkali metals Na and K: (i) an elemental abundance different from the commonly assumed solar value, (ii) masking by atmospheric hazes, (iii) in-situ photochemical reactions and/or (iv) condensation, e.g. on the planetary night side.

\citet{Lavvas2014} review processes affecting Na and K signals in hot Jupiter atmospheres and apply a photochemical model including in-situ Na and K reactions. Regarding the effect of elemental abundances, they suggested values up to 6$\times$ lower than solar for Na and K for e.g. HD209458 based on stellar metallicity measurements. Regarding the masking of spectral bands, they note that it is difficult to mask only the sodium line (and not the potassium line) since sodium is a strong absorber -  this could however be achieved in the presence of certain hazes although the size distribution of the haze particles would have to be consistent with the observed Rayleigh slope. Regarding in-situ photochemistry, two important species formed are:  XH (formed via:  X+H$_2$) and XOH (formed via X+H$_2$O) (where X=Na, K). Both XH and XOH can then undergo thermal decomposition to re-generate the alkali metal atoms (X). Uncertainties in the photochemistry include, for instance, reactions of potassium (e.g. here the three-body combination rates are not well known and one usually assumes the same coefficients as for sodium);  also at hot-Jupiter temperatures, excited states could become important, the photochemical responses of which are not well-defined. \citet{Lavvas2014} concluded that the uncertainties in photochemistry however, are small compared with the potential effect of masking by hazes. Regarding condensation of alkali metals on the night-side, more work is required on e.g. the 3D transport mechanisms across the terminator.

\section{Conclusions}

In this work, we have presented the transmission spectrum of the hot-Jupiter WASP-17b using the FORS2 instrument at ESO's VLT, in its multi-object spectroscopy mode.  Using a combination of light from multiple comparison stars, we obtained the broadband differential transit light curve of this planet, whereby the bulk orbital and physical parameters were derived.

From the broadband light curve, we obtained refined non-wavelength dependent transit parameters, consistent with previous analyses \citep{Anderson2011,Southworth2012,Bento2014}.  Through detailed study of parameter inference and correlations, we modelled all the obtained light curves using the quadratic limb darkening law for the host star.

Spectrophotometric light curves are analysed as independent GP noise models together with an analytical transit function, where strict, delta--function, priors for non-wavelength dependent parameters were assumed based on the broadband solution.  We explored the posterior distribution for the remaining free parameters, those being the scaled radius, the two coefficients of the limb darkening law, noise model parameters and three coefficients of a second order polynomial describing the colour--dependent out of transit flux variations, to quote the best fit solutions as a function of channel wavelength.  We take a non-parametric approach to modelling the time--correlated noise in the data, with time taken as the only input of our GP model \citep{Gibson2012b} in calculation of the covariance matrix.  This procedure was performed on two sets of light curves, where for the second set (CMC), we applied what is known as the common mode correction by removing the systematic noise common to all the wavelength channels.

Through comparison of transmission spectra (produced from both sets of light curves) with synthetic atmospheric models, we rule out a cloudy makeup of WASP-17b's upper atmosphere with high significance (>3$\sigma$).  From fitting a Rayleigh scattering slope we estimate an atmosphere with a mean molecular weight consistent with prevalence of H$_2$. Further observations with the 600B grism of FORS2, extending the spectrum towards the ultraviolet, will be required to confirm and quantify this aspect of the exo-atmosphere with higher precision. 

Additionally we looked closer at possibility of enhanced absorption towards the two main optical absorbers, sodium and potassium.  We do not detect a significant variation of the planetary radius at the sodium core with a 50\AA~bin, consistent with previous conclusions of \cite{Wood2011} and \cite{Zhou2012}.  Due to the low significance levels we not able to confirm nor rule out the presence of the pressure--broadened wings of the sodium absorption line.  Further, higher precision and resolution observations will be required to confirm this feature.  Similar conclusions are made for potassium, although we were not able to probe the absorption in the core of the line due to the telluric $O_2$ feature.  However, we do confirm the presence of the pressure--broadened wing of the potassium line with 3-$\sigma$ significance, which amounts to a significant detection.

Ultimately, our observations and analysis highlight the importance and capability of ground-based facilities in detecting and characterising exoplanetary atmospheres.  FORS2 will play an important role in those efforts, providing the wider community an essential outlet for followup of fascinating current and future targets.

\begin{acknowledgements}
We would like to thank the referee, Suzanne Aigrain, for great suggestions that hugely improved the quality of our analysis and conclusions. The science data together with all the calibration frames can be downloaded under the ESO archive number 095.C-0353(A).  We would also like to thank Aur\'elien Wyttenbach for useful discussions.  ES would like to acknowledge funding and support on this work from the ESO studentship programme. Part of the data analysis for this project was completed within the exoplanet atmosphere programme funded through the DAAD (project Nr. DAAD-2015-08). Some work by TJ was funded through the M\u{S}MT grant LG14033. MR is supported by STFC (ST/K502406/1) and the ERC project Exolights (617119). Sz. Cs. thanks the Hungarian OTKA for the Grant K113117. 
\end{acknowledgements}

\end{document}